\documentclass[preprint,showpacs,amsmath,amssymb]{revtex4}
\usepackage{epsfig}
\begin{document}
\def\bea{\begin{eqnarray}}
\def\eea{\end{eqnarray}}
\def\nn{\nonumber}
\newcommand{\snu}{\tilde \nu}
\newcommand{\sll}{\tilde{l}}
\newcommand{\asnu}{\bar{\tilde \nu}}
\newcommand{\stau}{\tilde \tau}
\newcommand{\dmsnu}{{\mbox{$\Delta m_{\tilde \nu}$}}}
\newcommand{\mt}{{\mbox{$\tilde m$}}}

\renewcommand\epsilon{\varepsilon}
\def\beq{\begin{equation}}
\def\eeq{\end{equation}}
\def\lla{\left\langle}
\def\rra{\right\rangle}
\def\za{\alpha}
\def\zb{\beta}
\def\lsim{\mathrel{\raise.3ex\hbox{$<$\kern-.75em\lower1ex\hbox{$\sim$}}} }
\def\gsim{\mathrel{\raise.3ex\hbox{$>$\kern-.75em\lower1ex\hbox{$\sim$}}} }
\newcommand{\Rbs}{\mbox{${{\scriptstyle \not}{\scriptscriptstyle R}}$}}

\draft


\title{Are deviation from  bi-maximal mixing and
none-zero $U_{e3}$ \\
related to  non-degeneracy of heavy Majorana neutrinos?
\bigskip\bigskip}


\thispagestyle{empty}
\author{ Sin~ Kyu~ Kang\footnote{ E-mail : skkang@phya.snu.ac.kr}}
\affiliation{ School of Physics, Seoul National University,
       Seoul 151-734, Korea }
\author{C. S. Kim\footnote{E-mail : cskim@yonsei.ac.kr} }
\affiliation{ Department of Physics, Yonsei University, Seoul
120-749, Korea
\bigskip\bigskip}

\begin{abstract}
\noindent {We propose a scenario that the mass splitting between
the first generation of the heavy Majorana neutrino and the other
two generations of degenerate heavy neutrinos in the seesaw
framework is responsible for the deviation of the solar mixing
angle from the maximal mixing, while keeping the maximal mixing
between the tau and muon neutrinos as it is. On top of the
scenario, we show that the tiny breaking of the degeneracy of the
two heavy Majorana neutrinos leads to the non-zero small mixing angle
$U_{e3}$ in the PMNS matrix and the little deviation of
the atmospheric neutrino mixing angle from the maximal mixing.
\bigskip\bigskip}
\end{abstract}

 \maketitle \thispagestyle{empty}
%

Thanks to enormous progress in solar, atmospheric and terrestrial
neutrino experiments, we have now the robust evidence for the
existence of neutrino oscillation which provides a window to
physics beyond the standard model (SM). Until now, while the
atmospheric neutrino deficit still points toward a maximal mixing
between the tau and muon neutrinos, however, the solar neutrino
problem favors a not-so-maximal mixing between the electron and
muon neutrinos. There have been many attempts to explain the
origin of the deviation of the solar mixing angle from the maximal
mixing. Surprisingly, it has recently been noted that the solar
neutrino mixing angle $\theta_{sol}$ required for a solution of
the solar neutrino problem and the Cabibbo angle $\theta_C$ reveal
a striking relation \cite{raidal}, $\theta_{sol}+\theta_C \simeq
\frac{\pi}{4}, $ which is satisfied by the experimental results
within a few percent accuracy,
$\theta_{sol}+\theta_{C}=45.4^{\circ}\pm 1.7^{\circ}$
\cite{SK2002,SNO,fits}. This quark-lepton complementarity (QLC)
relation has been simply interpreted  as an evidence for certain
quark-lepton symmetry or quark-lepton unification as shown in
Refs. \cite{raidal,smirnov,kkl}. But, it can be an accidental
phenomenon as pointed out in Ref. \cite{kkl,Jarlsk}. Thus, it is
worthwhile to find the possible alternatives to the grand
unification origin of the deviation of the solar mixing from the
maximal mixing.

In this letter, we propose a scenario that the mass splitting
between the first generation of the  heavy Majorana neutrino and
the other two generations of degenerate heavy neutrinos  in the
seesaw framework  is responsible for  the deviation of the solar
mixing angle from the maximal mixing, while keeping the maximal
mixing between the tau and muon neutrinos as it is. The maximal
atmospheric neutrino mixing and the smallness of $U_{e3}$ may be
the trace of the original ``bi-maximal'' mixing which is
presumably supposed to be achieved by some underlying flavor
symmetries, and thus the best possible approach to the problem is
to start in the limit of the maximal mixing with {\bf $U_{e3}=0$},
and understand how the deviation of the solar mixing from the
maximal is realized. In our scenario, the primitive ``bi-maximal"
neutrino mixing is generated only from the neutrino Dirac Yukawa
matrix by taking a diagonal form of three degenerate heavy
Majorana neutrinos in a basis where the charged lepton mass matrix
is real and diagonal. As will be shown, the deviation of the solar
mixing can then be generated from breakdown of the degeneracy of
the heavy Majorana neutrino masses between the first and the other
two generations. The main point in this scenario is that the
deviation  can be expressed in terms of the ratio between two
heavy Majorana neutrino masses. On top of the scenario, we will
also show that the tiny breaking of the degeneracy of the two
heavy Majorana neutrinos will lead to the small mixing angle
$\theta_{13}$ in the PMNS matrix and the very small deviation of
the atmospheric neutrino mixing angle from the maximal mixing.

Before proceeding to our scenario, we wish to motivate one
scheme that leads to exact ``bimaximal" mixing in the framework of
the seesaw mechanism. We study in a basis where the charged lepton
mass matrix is real and diagonal. The light neutrino mass matrix
$M_{\nu}$ diagonalized by $U_{\rm bimax}$ is given through the
seesaw mechanism by
\begin{eqnarray}
M_{\nu} & = & M^T_D~M_R^{-1}~M_D, \nonumber \\
& = & U_{\rm bimax}~ M^{\rm diag}_{\nu}~ U^T_{\rm bimax}~, \label{seesaw}
\end{eqnarray}
where $M_D=Y_D~v/\sqrt{2}$ with electroweak vacuum expectation
value $v$ and the neutrino Dirac Yukawa matrix $Y_D$,  and $M_R$
is a mass matrix of heavy Majorana neutrinos. The mixing matrix
$U_{\rm bimax}$ denotes the ``bi-maximal" mixing matrix
\cite{bimax}:
\begin{eqnarray}
U_{\rm bimax} &=& U_{23}\left(\frac{\pi}{4}\right)U_{12}\left(\frac{\pi}{4}\right) \nonumber \\
&=& \left(\begin{array}{ccc}
  \frac{1}{\sqrt{2}}& -\frac{1}{\sqrt{2}} & 0  \\
\frac{1}{2} & \frac{1}{2} &-\frac{1}{\sqrt{2}} \\
\frac{1}{2} & \frac{1}{2} & \frac{1}{\sqrt{2}}
 \end{array}\right). \label{bimax}
\end{eqnarray}
Then, the ``bi-maximal" mixing can be achieved by one of the three
possible ways as follows:
\begin{itemize}
\item{Taking  $Y_D$ diagonalized by $U_{\rm bimax}$ and
$M_R=M\cdot I$ with the identity matrix $I$ and a common mass
scale $M$.} \item{Taking $Y_{D}=y\cdot I$ and $M_R$ diagonalized
by $U_{\rm bimax}$.} \item{Taking ``bi-maximal" mixing pattern
from the combination of the nontrivial $Y_D$ and $M_R$.}
\end{itemize}
For the third case,  there may exist various origins of the
deviation of the solar mixing depending on possible combinations,
and some of which have been discussed before \cite{origin}. For
the other two cases, the modification of the trivial sectors
proportional to the unit matrix can be in charge for the origin of
the deviation from the maximal mixing. However, since the second
case may lead to the undesirable deviation of the atmospheric
mixing aside from the deviation of the solar mixing as one can
easily see, we only focus on the first case in this letter. In the
first case, the ``bi-maximal'' mixing can be achieved by taking
the symmetric matrix $Y_D$ with specific form. As an example, we
present a detailed model of $Y_D$ leading to the ``bi-maximal''
mixing, while keeping $M_R=M\cdot I$ based on the discrete
symmetry $A_4\otimes Z_2$ \cite{a4}. Let the three families of
leptons and singlet heavy neutrinos be denoted by $(\nu_i,l_i)_L$,
$l_{iR}$, $N_{iR}$ for $i=1,2,3$. In this convention,
$\bar{l}_{iL}l_{jR}$ and $\bar{\nu}_{iL}N_{jR}$ are Dirac mass
terms for charged leptons and neutrinos. Under the discrete
symmetry $A_4\otimes Z_2$, the 3 families of leptons transform as
$(\nu_i,l_i)_L\sim ({\bf 3,+}),~ N_{iR}\sim ({\bf 3, +}),
~l_{iR}\sim ({\bf 1},-), ({\bf 1}^{\prime},-),
({\bf 1}^{\prime \prime},-)
$.
We introduce Higgs scalar sectors consisted of seven
Higgs doublets $\Phi_i \sim ({\bf 1},-), ({\bf 1}^{\prime},-),
({\bf 1}^{\prime \prime},-), ~\phi \sim ({\bf 1,+}), ~\sigma_i \sim ({\bf 3,+})$.
{}From the assignment, the $A_4\otimes Z_2$ invariant Dirac
Yukawa interactions for charged lepton sector, $\overline{l}_{iL}l_{iR}\Phi_j$,
leads to a diagonal mass matrix with 3 independent entries each as shown
in Ref. \cite{chen}.
For the mass matrix of the heavy Majorana
neutrinos, we can take $MN_{iR}N_{iR}$ with common mass scale $M$
because of $A_4$ symmetry, {\it i.e.} ${\bf 3}\times {\bf 3} \sim
{\bf 1}$. The Dirac Yukawa matrix for the neutrino sector, which
is invariant under $A_4\otimes Z_2$ and diagonalized by the
``bi-maximal'' mixing matrix, can be obtained from the interaction
Lagrangian as follows:
\begin{eqnarray}
Y_D &=& h_1(\bar{\nu}_1N_1+\bar{\nu}_2N_2+\bar{\nu}_3N_3)\phi\nonumber \\
   &+& h_2(\bar{\nu}_1N_2\sigma_3+\bar{\nu}_2N_3\sigma_1+\bar{\nu}_3N_1\sigma_2) \nonumber \\
   &+& h_3(\bar{N}_1\nu_2\sigma_3+\bar{N}_2\nu_3\sigma_1+\bar{N}_3\nu_1\sigma_2)+h.c.
\label{eq3}
\end{eqnarray}
In order to achieve the symmetric form of the Dirac Yukawa matrix,
we require $h_2=h_3$.
The vacuum expectation values for the neutral components of Higgs
sector $\sigma^0_i$ can be determined by the Higgs potential
invariant under $A_4$,
\begin{eqnarray}
V &=& m^2\sigma^{\dagger}_i\sigma_i+\frac{1}{2}\lambda_1(\sigma^{\dagger}_i\sigma_i)^2 \nonumber \\
   &+&\lambda_2(\sigma^{\dagger}_1\sigma_1+\omega^2\sigma^{\dagger}_2\sigma_2
   +\omega\sigma^{\dagger}_3\sigma_3)(\sigma^{\dagger}_1\sigma_1+\omega\sigma^{\dagger}_2\sigma_2
   +\omega^2\sigma^{\dagger}_3\sigma_3)\nonumber \\
   &+&\lambda_3[(\sigma^{\dagger}_2\sigma_3)
   (\sigma^{\dagger}_3\sigma_2)+(\sigma^{\dagger}_3\sigma_1)
   (\sigma^{\dagger}_1\sigma_3)+(\sigma^{\dagger}_1\sigma_2)
   (\sigma^{\dagger}_2\sigma_1)]\nonumber \\
  &+& \left\{\frac{1}{2}\lambda_4[(\sigma^{\dagger}_2\sigma_3)^2
   +(\sigma^{\dagger}_3\sigma_1)^2
   +(\sigma^{\dagger}_1\sigma_2)^2 ]+h.c.\right\},
   \end{eqnarray}
where $\omega=e^{2\pi/3}$. Taking $<\sigma^0_1>=0$ and
$<\sigma^0_2>=<\sigma^0_3>=v$ with
$v=\sqrt{\frac{-m^2}{2\lambda_1+\lambda_2+\lambda_3+\lambda_4}}~$
as well as non-vanishing $<\phi^0>$ for the Higgs sector $\phi$,
we can achieve the final form of the Dirac Yukawa matrix
given as follows,
\begin{eqnarray}
Y_D=\left(\begin{array}{ccc}
 a & b & b \\
 b & a & 0 \\
 b & 0 & a \end{array}\right). \label{bimax}
\end{eqnarray}
Defining $Y^{\rm diag}_{D}=diag(x,y,z)$, the neutrino Dirac Yukawa
matrix $Y_D$ diagonalized by $U_{\rm bimax}$ is generally given in the
symmetric matrix form by
\begin{eqnarray}
Y_D=U_{\rm bimax}~Y^{\rm diag}_D~U^T_{\rm bimax}~. \label{dirac}
\end{eqnarray}
Here, we consider the case of nonzero values for $x$ and $y$,
which is crucial to our purpose.

In order to achieve the observed deviation of the solar neutrino
mixing from the maximal mixing, we take into account the mass
splitting between the first generation of the heavy Majorana
neutrino and the other two degenerate ones, for which the mass
matrix  is given by $M_R=M_R^{\rm diag}=(M_1,M_2,M_2)$, which
results from the breaking of $A_4$ in the heavy neutrino sector
and reflects separation of $N_{iR}\sim N_{1R}({\bf 1})\bigoplus
N_{(2,3)R}({\bf 2})$ under $S_3$ symmetry. Then, the light
neutrino mass matrix $M_{\nu}$ is presented as follows:
\begin{eqnarray}
M_{\nu} &=& U_{\rm bimax}
\left(\begin{array}{ccc}
 x & & \\
  & y & \\
  & & z \end{array} \right)U^T_{\rm bimax}
  \left(\begin{array}{ccc}
  M_1^{-1} & & \\
   & M_2^{-1} & \\
    & & M_2^{-1} \end{array} \right)U_{\rm bimax}
\left(\begin{array}{ccc}
 x & & \\
  & y & \\
  & & z \end{array} \right)U^T_{\rm bimax} \nonumber \\
&=&
U_{\rm bimax}
\left(\begin{array}{ccc}
 x & & \\
  & y & \\
  & & z \end{array} \right)U^T_{12}\left(\frac{\pi}{4}\right)
 \left(\begin{array}{ccc}
  M_1^{-1} & & \\
   & M_2^{-1} & \\
    & & M_2^{-1} \end{array} \right)U_{12}\left(\frac{\pi}{4}\right)
\left(\begin{array}{ccc}
 x & & \\
  & y & \\
  & & z \end{array} \right)U^T_{\rm bimax} \nonumber \\
  &=&
U_{\rm bimax} M^{\prime}_{\nu}
U^T_{\rm bimax}~,
\end{eqnarray}
where the mass matrix $M^{\prime}_{\nu}$ is given by
\begin{eqnarray}
M^{\prime}_{\nu}=
\left(\begin{array}{ccc}
\frac{x^2}{2M_1M_2}(M_1+M_2) & \frac{xy}{2M_1M_2}(M_1-M_2) & 0 \\
\frac{xy}{2M_1M_2}(M_1-M_2) & \frac{y^2}{2M_1M_2}(M_1+M_2) & 0 \\
0 & 0 & \frac{z^2}{M_2}
\end{array} \right) ~.
\end{eqnarray}
Then, the matrix $M^{\prime}_{\nu}$ can be diagonalized by
$U_{12}(\theta)$, and after diagonalizing $M^{\prime}_{\nu}$,
we can obtain the mixing angle $\theta$ and three neutrino mass
eigenvalues as follows:
\begin{eqnarray}
\tan 2\theta &=& \frac{2xy (M_2-M_1)}{(x^2-y^2)(M_1+M_2)}, \\
m_{\nu_1} &=& \frac{1}{2M_1M_2}[(c^2x^2+s^2y^2)(M_1+M_2)+2csxy(M_1-M_2)], \nonumber \\
m_{\nu_2} &=& \frac{1}{2M_1M_2}[(s^2x^2+c^2y^2)(M_1+M_2)-2csxy(M_1-M_2)], \\
m_{\nu_3} &=& \frac{z^2}{M_2}~, \nonumber
\end{eqnarray}
where $c=\cos\theta,~ s=\sin\theta$. Comparing the mixing matrix
$U_{12}(\theta)$ with $U_{12}(\pi/4)$ in $U_{\rm bimax}$, we can
get the solar mixing angle $\theta_{sol}$ which deviates as much
as the value of $\theta$ from the maximal mixing. Note that the
value of $\theta$ should be negative in order to achieve the
desirable deviation of the solar neutrino mixing. We can argue
that the generation of the mixing angle $\theta$ due to the
splitting between $M_1$ and $M_2$ in seesaw mechanism may be the
origin of the deviation of the solar mixing angle from the maximal
mixing in the case of nonzero $x$ and $y$. However, since we do
not have yet any information on the values of $M_1$ and $M_2$, we
cannot immediately test whether the difference between $M_1$ and
$M_2$ is really compatible with the deviation of the solar mixing
angle from the maximal mixing, but we can make numerical estimate
for the size of the ratio of $M_1$ to $M_2$, which accommodates
the deviation of the solar mixing based on the experimental
results for the neutrino oscillation. {}From the numerical
results, we can also predict the magnitude of the effective
Majorana neutrino mass $m_{ee}$, which is the neutrino-exchange
amplitude for the neutrinoless double beta decay.

For our purpose, let us define two parameters $\kappa$ and $\omega$ as follows:
\begin{eqnarray}
\kappa \equiv \frac{y}{x}~, ~~~~~\omega \equiv \frac{M_1}{M_2}~.
\label{ratios}
\end{eqnarray}
Then, the expressions for $\theta$ and $ m_{\nu_i}$ are given as follows,
\begin{eqnarray}
\tan 2\theta &=& \frac{2\kappa (1-\omega)}{(1-\kappa^2)(1+\omega)}, \label{tant}\\
m_{\nu_1} &=& \frac{x^2}{2M_1}[(c^2+s^2\kappa^2)(1+\omega)+2cs\kappa(\omega-1)], \nonumber \\
m_{\nu_2} &=& \frac{x^2}{2M_1}[(s^2+c^2\kappa^2)(1+\omega)-2cs\kappa(\omega-1)], \label{masseig} \\
m_{\nu_3} &=& \frac{z^2}{M_2}~. \nonumber
\end{eqnarray}
In addition, the effective Majorana neutrino mass $m_{ee}$
is presented by
\begin{eqnarray}
m_{ee}=\frac{x^2}{4M_1}[(1+\kappa)^2+\omega (1-\kappa)^2] \label{double}.
\end{eqnarray}
As shown in Eq. (\ref{tant}), the non-vanishing value of the
mixing angle $\theta$ can arise when $\omega$ is deviated from
one, which indicates the splitting between $M_1$ and $M_2$. In
fact, the present experimental results are not enough to determine
all the parameters introduced. But, if we fix one neutrino mass
eigenvalue by hand, we can determine several independent
parameters as well as the magnitude of $m_{ee}$ from Eqs.
(\ref{tant},\ref{masseig},\ref{double}). For our numerical
calculation, we set the parameter $\theta,~ \Delta m_{21}^2$ and
$\Delta m^2_{32}$ to be $13^{\circ},~ 8\times
10^{-5}~\mbox{eV}^2,~ 2.5\times 10^{-3}~\mbox{eV}^2$,
respectively. Those numbers correspond to the best fit values for
the measurements of the deviation of the solar mixing angle from
the maximal mixing, the mass-squared differences of the solar and
atmospheric neutrino oscillations, respectively. By fixing
$m_{\nu_1}$ as an input parameter, we can determine the parameter
set $(\kappa,\omega,\frac{x^2}{M_1},\frac{z^2}{M_2})$ for normal
hierarchy $m_{\nu_1}< m_{\nu_2} < m_{\nu_3} $ through the
relations (\ref{ratios},\ref{tant},\ref{masseig}).

\begin{table}[htb]
\caption{All numbers corresponding to the mass parameters are given in the unit eV for normal hierarchy.}
\begin{tabular}{cccccc}  \hline \hline
$m_{\nu_1}$(input) & $\kappa $ & $\omega$ & $\frac{x^2}{M_1}$ &
$\frac{z^2}{M_2}$ & $m_{ee}$\\
\hline
0.005 & 1.298 & 0.772 & 0.003 & 0.051 & 0.009 \\
0.01 & 1.118 & 0.897 & 0.006 & 0.052 & 0.013\\
0.05 & 1.006 & 0.994 & 0.025 & 0.071 & 0.051\\
0.1 & 1.002 & 0.998 & 0.050 & 0.112 & 0.101\\
 \hline \hline
\end{tabular}
\end{table}

In TABLE I, we present our numerical results for normal hierarchy.
{}From the TABLE I, we can see that the values of $\kappa$ and
$\omega$ approach to one as $m_{\nu_1}$ increases up to of order
0.1 eV, and one needs fine-tuning to obtain the parameter set
satisfying the relations above for the case of such a large
$m_{\nu_1}\sim 0.1$ eV. As $m_{\nu_1}$ goes down, the value of
$\kappa$ rapidly increases whereas that of $\omega$ decreases. We
can also predict the size of the amplitude of the neutrinoless
double beta decay $m_{ee}$ as a function of $m_{\nu_1}$, which is
presented in the last column of TABLE I. If the neutrinoless
double beta decay will be measured in near future, we will be able
to determine three neutrino mass eigenvalues and the parameters
introduced in Eqs. (\ref{tant},\ref{masseig},\ref{double}).
\begin{table}[htb]
\caption{All numbers corresponding to the mass parameters are given in the unit eV for inverted hierarchy.}
\begin{tabular}{cccccc}  \hline \hline
$m_{\nu_3}$(input) & $\kappa $ & $\omega$ & $\frac{x^2}{M_1}$ &
$\frac{z^2}{M_2}$ & $m_{ee}$\\
\hline
0.005 & 1.672 & 0.585 & 0.011 & 0.010 & 0.041 \\
0.01 & 1.569 & 0.630 & 0.013 & 0.013 & 0.042\\
0.05 & 1.137 & 0.881 & 0.029 & 0.051 & 0.066\\
0.1 & 1.044 & 0.959 & 0.052 & 0.100 & 0.109\\
 \hline \hline
\end{tabular}
\end{table}
For inverted hierarchy $m_{\nu_3}< m_{\nu_1}<m_{\nu_2}$, the
numerical results are presented in TABLE II. In this case,
contrary to the normal hierarchical case, we take $m_{\nu_3}$ as
an input.

Next, to generate non-vanishing $U_{e3}$, on top of the above
scenario, we consider an interesting possibility  that the
breaking of the degeneracy between the second and the third
generation masses in the heavy Majorana neutrino sector, {\it
i.e.}, $M_R=diag(M_1,M_2,M_3)$, can be an origin of the generation
of non-vanishing $U_{e3}$. We remark that the value of $U_{e3}$
goes to zero in the limit of $M_2=M_3$ in this scenario. The
effective light Majorana neutrino mass matrix is given by
\begin{eqnarray}
M_{\nu} &=& U_{\rm bimax}
\left(\begin{array}{ccc}
 x & & \\
  & y & \\
  & & z \end{array} \right)U^T_{\rm bimax}
  \left(\begin{array}{ccc}
  M_1^{-1} & & \\
   & M_2^{-1} & \\
    & & M_3^{-1} \end{array} \right)U_{\rm bimax}
\left(\begin{array}{ccc}
 x & & \\
  & y & \\
  & & z \end{array} \right)U^T_{\rm bimax} \\
    &=&
U_{\rm bimax}~ M^{\prime}_{\nu}~ U^T_{\rm bimax}~.
\end{eqnarray}
Assuming that the mass splitting between $M_2$ and $M_3$ is small enough to accommodate
the tiny $U_{e3}$, the mixing matrix,
which diagonalizes the neutrino mass matrix $M_{\nu}$,
can  be approximately given by
\begin{eqnarray}
U\simeq U_{23}\left(\frac{\pi}{4}\right)U_{12}\left(\frac{\pi}{4}\right)
\left( \begin{array}{ccc}
\cos \sigma & \sin \sigma & \delta \\
-\sin \sigma & \cos \sigma & \eta \\
-\delta & -\eta & 1 \end{array} \right),
\end{eqnarray}
where the mixing angle $\sigma$ corresponds to the (1,2) rotation of $2\times 2$
submatrix
of $M^{\prime}_{\nu}$.
The mixing angle $\sigma$ is presented by
\begin{eqnarray}
\tan \sigma\simeq \frac{2\kappa(1-\omega-\epsilon)}
{(1-\kappa^2)(1+\omega+\epsilon)},
\end{eqnarray}
where $\epsilon=M_1/M_3$, and $\omega,\kappa$ are given earlier.
This mixing angle $\sigma$ is responsible for the deviation of the
solar mixing angle from the maximal mixing. We note that
non-vanishing value of $\sigma$ is possible even when $\omega=1,
i.e. (M_1=M_2)$, but this case is undesirable because it leads to
negative $\sigma$ which {\it positively} contributes to
$\theta_{12}$. The mixing angle $\sigma$ is zero when
$\omega+\epsilon=1$, but it corresponds to the large hierarchy
among three heavy Majorana masses, which is far beyond our
purpose.
 The mixing elements
$\delta$ and $\eta$ are given by
\begin{eqnarray}
\delta &=& c_1\left(-\frac{1}{M_2}+\frac{1}{M_3}\right), \nonumber \\
\eta &=& c_2\left(-\frac{1}{M_2}+\frac{1}{M_3}\right),
\end{eqnarray}
where $c_1$ and $c_2$ are presented in terms of three light neutrino mass eigenvalues and
the parameters $\kappa,\omega,\epsilon$.
Then, the mixing element $U_{e3}$ and the deviation of the atmospheric mixing from
the maximal mixing are simply presented in terms of
$\sigma$ and $\eta$ as follows,
\begin{eqnarray}
|U_{e3}| &\simeq &  \frac{1}{2}|\delta - \eta|, \\
\delta \sin\theta_{23} &\simeq & \frac{1}{2}(\delta+\eta).
\end{eqnarray}
Imposing the bound on $|U_{e3}|$ of CHOOZ experiment,
$|U_{e3}|<0.2$, and the result of $\sin^2\theta_{23}$ from
atmospheric neutrino data,
$\sin^2\theta_{23}=0.44(1^{+0.41}_{-0.22})$ at $2\sigma$
\cite{fogli}, we can determine the allowed regions of the ratio
$M_2/M_3$. In TABLE III, we present the numerical results for the
ratio $M_2/M_3$ and the prediction for the bound on $|U_{e3}|$.
The second and third columns correspond to the normal hierarchical
case, whereas the fourth and fifth columns to the inverted
hierarchy. We find that the result for $\delta \sin^2\theta_{23}$
constrains $M_2/M_3$ more severely than the bound on $|U_{e3}|$
for $m_{\nu_{1(3)}}< 0.05$ eV. But for $m_{\nu_{1(3)}}\sim 0.1$
eV, both $\delta \sin^2\theta_{23}$ and $|U_{e3}|$ from neutrino
data severely constrain the allowed region of $M_2/M_3$. The
values in the columns for $|U_{e3}|$ indicate the predictions for
the upper bound. As shown in Table III, the allowed region for
$M_2/M_3$ gets narrowed as $m_{\nu_{1(3)}}$ increases, and it
becomes nearly one for $m_{\nu_{1(3)}}\geq 0.1$ eV. This implies
that such large values of $m_{\nu_{1(3)}}$ lead to moderately
degenerate light neutrino spectrum realized by almost degenerate
heavy Majorana neutrinos.

\begin{table}[htb]
\caption{The numerical results for the ratio $M_2/M_3$ and
the prediction for the bound on $|U_{e3}|$ for the normal hierarchical case and
the inverted hierarchical case.}
\begin{tabular}{c|cc|cc}  \hline \hline
$m_{\nu_{1(3)}}$(input) & $M_2/M_3 $ & $|U_{e3}|$ & $M_2/M_3$ & $|U_{e3}|$ \\
\hline
0.005 &$0.55\sim 1.29$ & $(<0.015)$ & $0.44\sim 1.36$ & $(< 0.025)$ \\
0.01 & $0.72\sim 1.16$ & $(<0.0007)$ & $0.59\sim 1.27 $ & $(< 0.015)$ \\
0.05 & $0.94\sim 1.04$ & $(<0.086)$ &  $0.93\sim 1.04$ &  $(< 0.083)$ \\
0.1 &  $0.98\sim 1.01$ &  & $0.98\sim 1.01 $ &  \\
 \hline \hline
\end{tabular}
\end{table}

Finally we note that there could be
radiative corrections to neutrino mass matrix which can lead to
some modification of our results. However, non-negligible
renormalization effects can be expected only in the case of
degenerate light neutrino spectrum. The numerical results for
$m_{\nu_1(3)}=0.1$ eV in the tables may be significantly modified
due to possible renormalization effects, but the detailed investigation on
the renormalization effects is not our main interest in this work
and we will leave it for the future work.

In summary, we have proposed a scenario that the mass splitting
between the first generation of the heavy Majorana neutrino and
the other two degenerate ones in the seesaw framework is
responsible for the deviation of the solar mixing angle from the
maximal mixing, while keeping the maximal mixing between the tau
and muon neutrinos as it is. Our scheme is based on the assumption
that nature presumably started with ``bi-maximal" neutrino mixing
and then it has been deviated somehow. We have considered the case
that the ``bi-maximal" mixing is achieved only from the neutrino
Dirac Yukawa matrix by taking a diagonal form of three degenerate
heavy Majorana neutrinos in a basis where the charged lepton mass
matrix is real and diagonal. Allowing the mass splitting between
the first and the other two generations of the heavy  Majorana
neutrinos, we could obtain the deviation of the solar mixing angle
from the maximal. In addition, we have also shown that the tiny
breaking of the degeneracy of the two heavy Majorana neutrinos
leads to the small mixing angle $\theta_{e3}$ in the PMNS matrix
and the very small deviation of the atmospheric neutrino mixing
angle from the maximal mixing.



\begin{center}
Acknowledgement
\end{center}
S.K.K. was supported in part by BK21 program of the Ministry of
Education in Korea and in part by KOSEF Grant No. R01-2003-000-10229-0.
C.S.K. was supported in part by  CHEP-SRC Program and
in part by the Korea Research Foundation Grant funded
by the Korean Government (MOEHRD) No. KRF-2005-070-C00030.


\end{document}